\documentclass[12pt,preprint]{aastex}

\usepackage{epsfig,rotating,amsmath}

\shortauthors{The Magellanic Stream}
\shorttitle{Putman et~al.}

\received{2002 April 6}
\begin{document}

\title{The Magellanic Stream, High-Velocity Clouds and the Sculptor Group}

\author{Mary E. Putman\altaffilmark{1},
        Lister Staveley-Smith\altaffilmark{2},
        Kenneth C. Freeman\altaffilmark{3},
        Brad K. Gibson\altaffilmark{4} and
        David G. Barnes\altaffilmark{4}}
\altaffiltext{1}{Center for Astrophysics and Space Astronomy, 
University of Colorado, Boulder, CO 80309-0389, USA; Hubble Fellow; mputman@casa.colorado.edu}
\altaffiltext{2}{Australia Telescope National Facility, 
                 CSIRO, P.O. Box 76, 
                 Epping, NSW 1710 Australia; Lister.Staveley-Smith@csiro.au}
\altaffiltext{3}{Research School of Astronomy \& Astrophysics,
                 Australian National University, Weston Creek P.O.,
                 Weston, ACT, Australia 2611; kcf@mso.anu.edu.au}
\altaffiltext{4}{Centre for Astrophysics \& Supercomputing,
                 Swinburne University, Mail \#31, P.O. Box 218,
                 Hawthorn, VIC, Australia 3122; bgibson@swin.edu.au, dbarnes@swin.edu.au}

\def\spose#1{\hbox to 0pt{#1\hss}}
\def\simlt{\mathrel{\spose{\lower 3pt\hbox{$\mathchar"218$}}
     \raise 2.0pt\hbox{$\mathchar"13C$}}}
\def\simgt{\mathrel{\spose{\lower 3pt\hbox{$\mathchar"218$}}
     \raise 2.0pt\hbox{$\mathchar"13E$}}}
\def\etal       {{\it et al.}}
\def\gtrapprox  {\;\lower 0.5ex\hbox{$\buildrel >\over \sim\ $}}
\def\lessapprox {\;\lower 0.5ex\hbox{$\buildrel < \over \sim\ $}}
\def\Msun       {${\rm M}_\odot$}
\def\deg        {$^\circ$}
\def\sigLL      {\sigma_{\scriptscriptstyle Lyc}}
\def\nH         {n_{\scriptscriptstyle H}}
\def\NH         {$\rm N_{\scriptscriptstyle H}$}
\def\HI         {\ion{H}{1}}
\def\Ha         {${\rm H}\alpha$}
\def\eg         {{\it e.g.,\ }}
\def\ie         {{\it i.e.,\ }}
\def\cf         {{\it cf.\ }}
\def\qv         {{\it q.v.,\ }}
\def\kms        {km~s$^{-1}$}
\def\cmmsq      {cm$^{-2}$}

\def\plotfiddle#1#2#3#4#5#6#7{\centering \leavevmode
\vbox to#2{\rule{0pt}{#2}}
\includegraphics{#1}}
\begin{abstract}
  The Magellanic Stream is a $100\arcdeg \times 10\arcdeg$ filament of
  gas which lies within the Galactic halo and contains $\sim 2 \times
  10^8$ \Msun\ of neutral hydrogen.  In this paper we present data
  from the HI Parkes All Sky Survey (HIPASS) in the first complete survey of the entire Magellanic
  Stream and its surroundings. We also present a summary of the
  reprocessing techniques used to recover large-scale structure in the
  Stream. The substantial improvement in spatial resolution and
  angular coverage compared to previous surveys reveals a variety of
  prominent features, including: bifurcation along the main Stream filament;
  dense, isolated clouds which follow the entire length of the Stream;
  head-tail structures; and a complex filamentary web at the head of
  the Stream where gas is being freshly stripped away from the Small
  Magellanic Cloud and the Bridge.  Debris which appears to be of
  Magellanic origin extends out to 20\arcdeg\ from the main Stream filaments.  
  The Magellanic Stream has
  a velocity gradient of 700 \kms\ from the Clouds to the tail of the
  Stream, $\sim 390$ \kms\ greater than that due to Galactic
  rotation alone, therefore implying a non-circular orbit. The dual
  filaments comprising the Stream are likely to be relics from gas
  stripped separately from the Magellanic Bridge and the SMC. This
  implies: (a) the Bridge is somewhat older than conventionally
  assumed; and (b) the Clouds have been bound together for at least
  one or two orbits. The transverse velocity gradient of the Stream
  also appears to support long-term binary motion of the Clouds. A significant
  number of the most elongated cataloged Stream clouds (containing $\sim 1$\% of
  the Stream mass) have position angles aligned along the Stream. This
  suggests the presence of shearing motions within the Stream, arising
  from tidal forces or interaction with the tenuous Galactic halo.  As
  previously noted, clouds within one region of the Stream, along the
  sightline to the less distant half (southern half on the sky) of the Sculptor Group, 
  show anomalous properties.  There are more clouds along this sightline than any other part
of the Stream and their velocity distribution significantly deviates from the gradient along the Stream.
We argue that this deviation could be due to a
  combination of halo material, and not 
  to distant Sculptor clouds based on a spatial and kinematic comparison between the Sculptor
Group galaxies and the anomalous clouds, and the lack of cloud detection in the northern half
of the group.  This result has significant
  implications for the hypothesis that there might exist distant,
  massive HVCs within the Local Group.  Cataloged clouds within the Magellanic
  Stream do not have a preferred scale size.  Their mass spectrum
  $f(M_{\rm HI})\propto M_{\rm HI}^{-2.0}$ and column density spectrum
  $f(N_{HI})\propto N_{\rm HI}^{-2.8}$ are steep compared with
  Ly$\alpha$ absorbers and galaxies, and similar to the anomalous clouds
along the Sculptor Group sightline.
\end{abstract}


\section{Introduction}
\label{introduction}

Many searches have been made for streams of halo material; the remnants of
Galactic satellites which are
responsible for building up the Milky Way (e.g. Newberg et al. 2002; Morrison \etal 2000;
Majewski \etal 2000; Lynden-Bell \& Lynden-Bell 1995).  These
searches have concentrated on the stellar halo, but neutral hydrogen is 
also a key tracer of galaxy formation and
destruction in the low redshift universe (e.g. Ryder et al. 2001; Smith 2000; Hibbard \& Yun 1999; Yun, Ho \& Lo
1994).  Our Galaxy is a prime example of this, with neutral hydrogen streams and
their remnants tracing the more recent aspects of the Milky Way's
formation and evolution.  
The most famous Galactic halo HI stream is the Magellanic Stream.
Discovered 30 yrs ago (Wannier \& Wrixon 1972; Mathewson
\etal 1974), this complex arc of neutral hydrogen starts
from the Magellanic Clouds and continues for over 100$^{\circ}$ through
the South Galactic Pole.  It has been created through the interaction
of our Galaxy with the Magellanic Clouds and may represent a recent
example of the accretion and merging process which created the Milky
Way.   The finding of a leading  stream of material (the Leading Arm)
indicates that the dominant mechanism responsible for forming the
Stream is tidal (Putman \etal 1998), however it remains unclear how
much the passage through the Galaxy's diffuse corona or extended disk
may have shaped this feature.

The accretion and merging of HI clouds may be occurring throughout the
Local Group and there has been a great deal of controversy as to
whether some of the high-velocity clouds (HVCs) represent the leftover Local Group
building blocks at average distances of 700 kpc from the Milky Way
 (e.g. Blitz \etal 1999; Braun \& Burton 1999).  Many of the clouds would have HI masses 
of a few times $10^7$ \Msun\ at Local Group distances; however, if only the compact,
isolated HVCs are at large distances, the HI mass of a typical intra-Local Group
cloud would only be a
few times $10^6$ \Msun\ (Putman \etal 2002).  The detection of star-free intergalactic
HI clouds in nearby
groups which are kinematically similar to the Local Group would greatly
support a Local Group origin for HVCs.
Deep HI surveys of nearby groups have detected 0 intergalactic HI clouds to masses of a 
few $\times 10^7$ \Msun~(e.g. Zwaan 2001; Banks \etal
1999).  Recently, de Blok et al. (2001) surveyed 2\% of the total area of the
  Sculptor and Centaurus A groups to HI masses of a few times $10^6$ \Msun\ and also
  found 0 free-floating HI clouds.  Considering the mass limitations of the
  large scale surveys and the limited area covered in the
  Sculptor and Centaurus A group survey, the proposal of compact HVCs being
  scattered throughout the Local Group remains a possibility.  

The Local Group is unvirialized and is falling together for the first time (Schmoldt \& Saha 1998). 
A nearby group which may be kinematically similar to the Local Group is the
Sculptor Group.  The Sculptor Group is also unvirialized and
appears to form part of a large continuous filament of galaxies which includes the
Local Group (Cote \etal 1997; Jerjen \etal 1998).
There is growing evidence that intergalactic clouds (Ly$\alpha$
absorbers) trace filaments and clusters of galaxies (Penton, Shull \&
Stocke 2000), and the unsettled environment of the Sculptor-Local
Group cloud could be the ideal environment to find these intergalactic
clouds in both absorption and emission. 

The Sculptor Group falls along the same sightline as a section of the
Magellanic Stream.  This, together with the low velocities of the
Sculptor Group galaxies (down to 70 \kms), make it very difficult to
distinguish how much of the high-velocity neutral hydrogen is directly
related to the Magellanic Stream and how much may be intra-Sculptor
Group/Local Group material.  Mathewson \etal~ (1975) was the first to
argue that some of the HI clouds along the sightline  to the
Magellanic Stream do not seem to fit into the normal velocity
distribution of the Stream and may be associated with Sculptor Group
galaxies.
This was countered by Haynes \& Roberts (1979) and Haynes (1979) who
argued it is merely a spatial coincidence.
The origin of these clouds was ultimately left unanswered, with the
possibility of a detection of intergalactic HI clouds remaining
alluring.

Uncovering the Magellanic Stream and its surroundings is crucial to
the understanding of the formation and evolution of not only the Milky
Way, but the entire Local Group, yet until the present paper, the
neutral hydrogen data for the entire Stream and its surroundings
remained severely under-sampled (e.g. Mathewson \etal 1974; Bajaja
\etal 1985).  Sections of the Stream have been observed at higher
spatial or velocity resolution (e.g. Cohen 1982; Haynes 1979; Morras
1983, 1985; Wayte 1989),  but these observations focused on small
regions within the main filament of the Stream originally presented by
Mathewson \etal.  Here we present HI maps of the entire
Magellanic Stream and its surroundings at 15.5$^{\prime}$ resolution
(equivalent to 250 pc at 55 kpc) using data from the HI Parkes All
Sky Survey (HIPASS; Barnes \etal~ 2001).
Though a tidal origin may be the primary one for the Stream, there is
clearly a complex history behind the formation of this feature.  The
large area of the survey also provides new insight into the origin of
the clouds along the Sculptor Group sightline.  We begin this paper by
describing the HIPASS observations and the HVC data reduction technique (known
as {\sc minmed5}) and
subsequently describe the spatial and kinematic HI structure of the
Stream and the high-velocity clouds along the Sculptor Group
sightline.  We then go on to discuss the origin of the
Stream and the anomalous clouds along the Sculptor Group sightline.

\section{Observations and Data Reduction}

This paper uses data from the \ion{H}{1} Parkes All-Sky Survey
(HIPASS).  HIPASS is a blind survey
for \ion{H}{1} and, including the northern extension, covers the sky
south of Decl. 25\arcdeg.  HIPASS was completed with the 64-m Parkes radio
telescope installed with a multibeam
receiver, or focal-plane array of 13 beams set in a hexagonal grid
(Staveley-Smith \etal\ 1996).  HIPASS observations consist of active
scans of the sky in strips 8\arcdeg\ long in Decl., with
 Nyquist sampling between scans.  The multibeam correlator has a total bandwidth of 64 MHz with 1024
channels for each polarisation and beam. The velocity range covered by
HIPASS is approximately $-1200$ to 12700 \kms, though the
re-processing described here only considered the velocity range $-700$ to 1000 \kms\ 
(LSR reference frame). The corresponding channel spacing is 13.2 \kms\ 
and the spectral resolution, after Hanning smoothing, is 26.4 \kms.  
See Barnes \etal\ (2001) for further details on HIPASS.

HIPASS was originally designed to detect discrete \HI\ sources
(i.e. galaxies) and therefore the data reduction technique was
optimized for such sources.  This normal HIPASS
reduction technique imposes a severe spatial filter on the sky (most
galaxies being compact relative to the Magellanic Stream), and so to
recover the large scale structure of the clouds and obtain accurate fluxes 
the data was re-processed with a different algorithm.  The normal HIPASS
reduction calculates the bandpass correction through a median of the reference spectra taken $\pm
2\arcdeg$ from the target spectrum, filtering
out emission which extends over angular scales greater than 2\arcdeg\ 
in Decl..  The modified version of the bandpass algorithm, the {\sc
  minmed5} method, utilizes the entire 8\arcdeg\ scan to recover large scale
\HI\ emission.  For each channel, beam and
polarization, {\sc minmed5} breaks each scan into five sections
(1.6\arcdeg\ long in Decl.), finds the median flux density for each
section, and uses the minimum of the five median values to form the
template bandpass for the entire 8\arcdeg\ scan.  {\sc minmed5}
greatly increases the sensitivity to large-scale structure and reduces
spatial sidelobes. A source needs to extend over greater than $\sim$
6\arcdeg\ in Decl.  before its flux is not measured well by {\sc
  minmed5}.  Since the bandpass can vary in time, {\sc minmed5} 
does not have the stability of the original HIPASS bandpass removal;
however, the only effect of this is a slight striping in the final
cubes along the RA axis.  Other steps in the revised reduction pipeline include
a median baseline fit, Hanning smoothing and a velocity conversion to
values relative to the local standard of rest.  The Hanning smooth
greatly reduces spectral ringing near bright Galactic \ion{H}{1} where
HVCs were previously completely hidden in the HIPASS-processed data.  The hanning smooth
may also make the detection of very narrow line-width sources less likely.

Figures 1 and 2 depict the difference in channel maps and integrated
intensity maps of HIPASS data containing extended emission reduced with
the standard HIPASS reduction method and {\sc minmed5}.  The
standard method clearly filters out emission, resulting in negative
sidelobes. {\sc minmed5} greatly improves the quality of intensity
(Figure 1) and integrated intensity (Figure 2) maps, and the amount of
negative flux is closer to zero.  An example of
the difference in terms of emission recovered per channel is
represented in the spectra of Figure 3 which again shows the original
HIPASS reduction (dashed line) versus {\sc minmed5} (solid line), for the 
sightline towards Fairall 9 through the Stream.  The emission lost in the individual channels
is apparent.  With 
{\sc minmed5} we obtain a column density of $8.0 \times 10^{19}$ cm$^{-2}$ 
for the Fairall 9 sightline through the Stream, which is within 15\% of the value obtained 
from the Parkes Multibeam narrow-band
data (Gibson \etal\ 2000).
Comparison of the {\sc minmed5} reduced HIPASS data to the Leiden-Dwingeloo Survey indicates an average difference
of 50\% for $N_{HI}$ and 25\% for $T_{max}$.  This type of difference
is expected considering the different resolutions of the surveys and the
clumpiness of HVCs on arcminute scales (e.g. Wakker \etal\ 2001).

Some regions near the Galactic Plane and the Small Magellanic Cloud
(SMC) still show sidelobe artifacts and loss of flux density with
{\sc minmed5} due to the emission filling the entire 8\deg\ scan. 
In these regions, a total-power \ion{H}{1} survey such
as the Leiden-Dwingeloo Survey (Hartmann \& Burton 1997) or the Parkes
narrow-band survey (Br\"uns, Kerp \& Staveley-Smith 2000) gives more
accurate flux densities and temperatures. For example, Stanimirovic
\etal\ (1999) quote a maximum column density in the SMC of $1.14\times
10^{22}$ atoms \cmmsq. The corresponding column density in the present
data is $7.4\times 10^{21}$ atoms \cmmsq, equivalent to a loss of 35\%
in flux density.

Since individual HIPASS scans are 8\deg~ long in Decl. and only
1.7\deg~ wide in RA, they need to be combined to study extended
features such as high-velocity clouds.  The gridding process also
incorporates overlapping spectra into a single cube, as each scan overlaps by
$\sim$ 1\deg~ and the survey re-observes each sky point 5
times (called the 'a' through 'e' scans).  The
standard HIPASS gridding process uses a weighted median of all pixels
within a $6^{\prime}$ radius to determine the flux for a specific
position, as this method is simple and robust against interference.  
The weighting over-corrects the fluxes for extended sources and so a simple median
was used for the HVC survey data (i.e. no re-normalization of the data was
done).
Again, see Barnes \etal~~2001 for a full description of the gridding algorithm
and the flux density
correction for objects less than $30^{\prime}$ in size.  The gridding process increases the beamwidth
from 14\farcm3 to $\sim$ 15\farcm5.  

The HVC HIPASS cubes are 24\deg\ $\times$
24\deg, and were subsequently mosaiced together into Zenithal Equal Area projection centered on the South
Galactic Pole to create the maps presented here.  The pixel size in the gridded cubes is 4\arcmin, but 
this was increased to 8\arcmin\ in the final mosaic in
order to reduce the size of the data set. Each pixel in the
final mosaic was a weighted sum of pixels in the input cubes.
Normally, the weight for each pixel in the input cubes was made equal
to the number of HIPASS spectra contributing to that pixel. However,
weights less than 25 were set to zero in order to increase the
immunity of the final images to interference.  The rms noise of the
final mosaic varies somewhat with velocity and position.  Away from obvious
regions of emission, the rms is 8--10 mJy/beam, corresponding to an rms
brightness temperature sensitivity of 7--8 mK. A 5-$\sigma$ detection of
a line of width 35 \kms\ therefore corresponds to a HI column density
detection level of $2.2 \times 10^{18}$ cm$^{-2}$.  The northern
(Decl. $>+2\arcdeg$) data presented here are of lower sensitivity, as only $\sim 20$\% of the total integration time was
available at the time the mosaic was made. The rms for this data is $\sim$14 mK.  The key parameters of the
data are summarized in Table 1.

\section{HI Distribution \& Kinematics}

\subsection{The Magellanic Stream}

The \HI\ column density image of the Magellanic Stream and the Clouds
is shown in Fig.4.  An annotated version of
this image is given in Galactic coordinates in Fig.5, showing the various recognised
components of the Magellanic System. Fig.5 is also overlayed with a
set of markers which show the approximate Magellanic Longitude for
different parts of the System.  The Magellanic coordinate system has
been previously described by Wakker (2001) and was chosen as a simple
way to have the Magellanic Stream lie at approximately the equator.  
The north pole of the Magellanic coordinate system is
defined to lie at Galactic coordinates $l=180$\deg, $b=0$\deg.  The
equator of the Magellanic coordinate system
passes through the South Galactic Pole and through the Galactic
equator at Galactic coordinates $l=90\arcdeg$ and 270\arcdeg.  The Stream is roughly
parallel to the equator of the Magellanic coordinate system, but generally
lies between Magellanic Latitudes, $0\arcdeg < B_M < -18\arcdeg$.
Magellanic Longitude, $L_M = 0\arcdeg$ is defined to correspond to the
centre of the LMC. This is not the same coordinate system used by Wannier \&
Wrixon (1972). Individual channel maps of the region shown in Fig.5 are shown in
the LSR velocity frame in Fig.6.  The velocity field, also in the LSR frame, is shown in
Fig.7.  Note that for Dec $> +2$\deg, or $L_M \approx 240$\deg\ to 270\deg,
the data is predominantly the incomplete northern HIPASS data, with
a 5$\sigma$ sensitivity of $\sim 0.07$ K instead of 0.035 K.

The \HI\ masses and mean column densities of the various components
annotated in Fig.5 are listed in Table~2. For masses, we use distances
to the LMC, Bridge and SMC of 50, 55 and 60 kpc, respectively. The
Stream distance is much more problematic. We normalise it at 55 kpc
which may be appropriate for the head of the Stream, but is possibly
wrong by a factor of 2 at the tail.  Mean column densities are
calculated from areas with emission above the sensitivity quoted in
Table~1.  Our \HI\ mass estimates for the LMC and SMC alone are
$2.9\times10^8$ M$_{\odot}$ and $3.4\times10^8$ M$_{\odot}$,
respectively. As expected, these values are slightly below other
determinations. For the LMC, Luks \& Rohlfs (1992) quote $3.1\times10^8$
M$_{\odot}$ and Staveley-Smith \etal\ (2002) quote $4.8\times10^8$
M$_{\odot}$. For the SMC, Stanimirovic et al. (1999) quote
$3.8\times10^8$ M$_{\odot}$ (excluding the self-absorption
correction).

The total amount of neutral hydrogen shown in Fig.~\ref{fig5},
including the Magellanic Clouds, but excluding the Leading Arm, 
and accounting for the
distances of the various components, is $9.0\times10^8$ M$_{\odot}$ (see Table 2).  The Leading Arm at
$b<0$\deg\ has a mass of around 10\% that of the Stream or $\sim 2 \times 10^7$ M$_{\odot
}$ (Putman et al. 1998).  For the Stream itself, if we
exclude the Magellanic Bridge and the velocity range $\pm 20$ \kms\ 
which is confused with Galactic emission, the integrated \HI\ mass is
$1.9\times10^8$ M$_{\odot}$ (again for an assumed distance of 55 kpc).
As seen in Fig.~\ref{fig6}, emission from the Stream is reasonably well-defined
near 0 \kms. An attempt to isolate Stream emission from Galactic
emission raises our estimate for the mass of the Stream to
$2.1\times10^8$ M$_{\odot}$ (see Table~\ref{tbl2}).  Wakker (2001), by
comparison, estimates a somewhat lower Stream mass at 55 kpc of $1.2 \times
10^8$ M$_{\odot}$ (based on the data of Hulsbosch \& Wakker (1988) and Morras 
et al (2000)), however he has also quoted $6 \times 10^8$ M$_{\odot}$ (Wakker \& van Woerden
1991), representing the ambiguity in defining the Stream.
Most of the Stream's mass lies at its head, close to the Clouds and
the Bridge (between the LMC/Bridge/SMC and MS~I in Fig.~\ref{fig5}). This region,
which predominantly lies in the Magellanic longitude range
$L_M=330$\deg\ to 350\deg\ or the Galactic latitude range
$b=-60$\deg\ to $-45$\deg\ (see Fig.~\ref{fig5}), contains approximately
half the total mass of the Stream, $\sim 1.1\times10^8$ M$_{\odot}$.
Much of this \HI\ resides within numerous and complex filaments which
are discussed below.

The Magellanic Stream has traditionally been represented as a single
continuous filament with several concentrations along its length (e.g. Mathewson \etal~ 1974), but the HIPASS data
clearly show a more complex structure primarily made up of two distinct and parallel
filaments.  This bifurcation has previously been noted by Cohen (1982)
and Morras (1983, 1985).  The dual filaments run parallel to each
other for the length of the Stream, but appear to merge three or more
times. At these points, for example $(\ell, b) = (40\arcdeg,
-82\arcdeg)$ (MS~III) and $(74\arcdeg, -68\arcdeg)$ (MS~IV) there are
dense concentrations of gas.  The two filaments are most evident at Magellanic
longitudes $L_M<340\arcdeg$ due to the complexity at the head of the
Stream. The filaments become thinner and appear to get closer together
as MS~IV (at $L_M=282\arcdeg$) is approached, somewhat suggestive of an
increasing distance (a factor of $\sim2$ would be needed). However,
beyond this the Stream again appears to fan out and become more
chaotic. The concentration of gas at MS~IV looks like a bow-shock,
suggestive that interaction with halo gas may be responsible for the
appearance of the Stream in this region. This is further discussed
below.

The two filaments of the Stream give the impression of
twisting about each other in a double helix, akin to a DNA molecule.
This appears to represent the pseudo-binary motion of the LMC and the SMC,
the progenitors. In the models considered by Gardiner et al. (1994), 
the orbital timescale of the SMC about the LMC is $\sim 0.9$ Gyr. 
This would imply $\sim 1.7$ orbits since the likely epoch of the creation 
of the Stream $\sim 1.5$ Gyr ago (Gardiner \& Noguchi 1996). This 
corresponds to $\sim 3-4$ crossings of the orbital plane defined 
by the barycenter of the Clouds around the Galaxy,
consistent with our observations. This implies that the
bifurcation of gas in the Stream is intrinsic and
reflects the different origin of the filaments. It is unnecessary
to invoke large-scale shocks, for example, to create the
dense concentrations of gas discussed in the previous paragraph. We argue that the 
two streams most plausibly arise from gas stripped from the Bridge and 
the SMC. The separation
of these components is similar to the maximum projected separation of 
$\sim 9$ kpc between the dual filaments. Moreover, the present-day structure
of the head of the Stream (see below) is also consistent with such a scenario.
A corollary of this scenario is that the Magellanic Clouds must have
shared a common gaseous envelope prior to the more recent ($\sim 0.2$ Gyr) 
encounter between them.

The gradual decrease in column density noted by Mathewson and the
natural separation of the Stream into several discrete clumps is not
immediately apparent in Fig.4, being hidden by the complexity of the
higher spatial resolution HIPASS data.  However, if the main filament
of the Stream is isolated and broken up into $\sim10\arcdeg$ sections,
the mean column density gradually decreases for $L_M<328\arcdeg$ (see
Fig.5 and Table 2).  MS~I-III have the highest mean column densities
$<N_{HI}>\approx 3.4\times 10^{19}$ cm$^{-2}$. The northern tail (MS~VI)
has the lowest mean column density at $3.6\times 10^{18}$ cm$^{-2}$.
As mentioned previously, the Stream's velocity coincides with local
Galactic emission at $(\ell, b) \approx (315\arcdeg, -80\arcdeg)$ and
some detailed information is lost. 
An attempt to correct for the confusion with Galactic emission has been made in Table~2.

The beginning, or head, of the Stream is characterised by a network of
numerous filaments and clumps. The head appears to emanate from the
northern side of the Magellanic Bridge and SMC at velocities between 
$v_{\rm LSR}= 90-240$ \kms~ (see
Figures 5 to 7).  The velocity of the beginning of each filament in
this region is similar to the corresponding Bridge velocity. That is,
there is a general velocity gradient across the head of the Stream
which coincides with the general velocity gradient seen in the Bridge
between the LMC at $\sim 300$ \kms\ and the SMC at $\sim 100$ \kms.  A
small amount of Stream material near the northern edge of the LMC has
a high positive velocity of 390 \kms.  This filament, at
$\ell=272\arcdeg, b=-40\arcdeg$ (see Fig.5), extends almost
perpendicular to the majority of filaments and appears to be of a
different nature.  The filaments in the head of the Stream vary in
length and width but are mostly characterised in being extremely
clumpy.  The high column density clumps with $N_{HI}\approx 1 - 2.5
\times 10^{20}$ cm$^{-2}$ are joined by more diffuse filaments with
$N_{HI}\approx 1.5 - 3 \times 10^{19}$ cm$^{-2}$.  Most of the
filaments project straight from the Bridge and SMC, but there are also
tenuous connections between the filaments themselves which make the
structure appear web-like.

With the exception of the two main filaments which continue on to form
what is canonically known as the Magellanic Stream, the filaments
making up the head of the Stream gradually decrease in column density
and disperse by $L_M < 330\arcdeg$ and $v_{\rm LSR}< 85$ \kms.  There is a
slight break in the continuity of the Stream's emission at
approximately the same position as the end of these filaments (see
Fig.4).  After this $\sim 4\arcdeg$ break, two of the larger filaments
continue northward.  Though this break is not an artifact, the slight
break at $\ell\approx 300\arcdeg$, $b\approx -54\arcdeg$ is partially
due to {\sc minmed5} and the emission completely filling an 8\arcdeg\ scan.

Following its chaotic beginnings near the Clouds, the Stream becomes
more confined as the Magellanic longitude decreases. The central
filaments continue for another $\sim 80$\deg, passing through
$v_{\rm LSR}$ = 0 \kms\ at $\ell=310$\deg, $b=-78$\deg\ (RA
$00^h44^m$, Decl. $-39$\deg) and proceeding to lower $L_M$'s at increasingly
negative velocities.  The velocity gradient along the Stream is
striking and much of it is clearly a reflex motion due to solar motion
around the Galactic Center. The Stream begins near the Clouds at
$v_{\rm LSR} \approx 250$ \kms\ and extends to $-450$ \kms\ at
its tail, an overall range of 700 \kms. In terms of a Galactic
reference frame, it begins at $v_{\rm GSR} \approx 100$ \kms\ and extends
to $-290$ \kms.  This is still substantial ($\sim 390$ \kms) and
indicates there are non-circular motions present.
Mathewson et al. (1974) point out that this variation in radial velocity is what is
expected if the gas is in a Keplerian orbit and seen from a focus.

There is also a transverse gradient to the main filament of the Stream of
the order of 5.6 \kms\ deg$^{-1}$ (Cohen 1982).
The velocity resolution of HIPASS is not ideal for studying this, but a
clear transverse gradient is apparent at $L_M \approx 300\arcdeg$ in the sense that 
velocities decrease as Magellanic Latitude increases. The gradient is of 
similar magnitude, but reversed in direction to the current velocity gradient 
in the Cloud/Inter-Cloud region. This may again reflect the 
orbital history of the Clouds.

 At $L_M<270$\deg, (north of Decl. $\sim 0$\deg) the 
two main filaments fan out by $\sim 10$\deg.  They break up
into a network of clumps and thin, diffuse connections, somewhat
reminiscent of the structure seen at the head of the Stream, but
at much lower column densities.
Many of the clumps at the northern tail have a head-tail structure
with the tails pointing along the major axis 
of the Stream.
The clumps remain at relatively high column densities ($N_{HI}\approx 10^{19}$
\cmmsq) until $L_M \approx 255$\deg\ where the Stream becomes
a series of diffuse filaments with peak column density concentrations
on the order of a few times $10^{18}$ \cmmsq. This may represent the
formation point of the Stream, or the point at which gas stripping and
evaporation has overwhelmed the Stream.

Dense clumps of \HI, which follow the central filaments of the Stream
in position and velocity, are a ubiquitous feature as shown in Figures~\ref{fig5} 
to~\ref{fig7}.  At the same Magellanic longitude along the main filaments, 
the clumps can be found at higher and lower latitudes (by up to 10\deg)
and at higher and lower velocities.  Many of the clumps, both in and about 
the Stream, show head-tail structures (i.e. a dense core with diffuse 
tail structures in both position and velocity). This is especially true at 
the head of the Stream.  Often, but not always, the tails point away from 
the Clouds.  A few prime examples of such features are shown in Figure~\ref{fig8}.   
The tail shown in the top panel points away from the Clouds. The tails in
the bottom panel, which are clouds along the
Sculptor Group sightline (see the next section), do not.
The column density distribution along the clumps is characterised by a 
sharp cutoff at the high column density end, and a gradual dispersal into 
a tail which is typically about twice the length of the head.
Examination of a small sample of dense clumps indicates that the 
tail typically contains half the \HI\ mass present in the 
head.  Some of the clumps at positive velocities between $L_M = 295$\deg\ 
to 310\deg\ in Fig.~\ref{fig6} and~\ref{fig7}
are actually galaxies of the Sculptor Group.  
The distribution of individual clouds along the Magellanic Stream 
is investigated in more detail in the next section.

\subsection{Distribution of Clouds}

We now examine the properties and distribution of the clouds
within and around the Magellanic Stream as cataloged by Putman et
al. (2002).  The clouds north of Decl $+2\arcdeg$ are not
cataloged due to the lower sensitivity of this data.  
Only clouds with 
velocities in the range $-500 < v_{\rm LSR} < -80$ \kms\ or 
$80 < v_{\rm LSR} < 500$ \kms\ are considered here (there were no clouds detected at
higher velocities),
and all known galaxies were excised. 
The steps used to catalog the clouds can be summarised as follows:

\begin{enumerate}
\item All pixels with $T_B > 6$ mK were examined, and each was assigned to a 
local maximum with which it is connected (spatially and in velocity). 
\item Only those local maxima with $T_B$(max)$ > 12$ mK are kept.
\item Adjacent local maxima were merged into a single cloud if the
  brightest enclosing contour for the maxima has $T_B(\circ) > 80$ mK (i.e.
  there exists a bright connection between the maxima) or if $T_B(\circ)> 0.4
  T_B$(max) (i.e. the contrast between the local maxima is small).
\item Finally, the merged maxima were only labelled as clouds if
  $T_B$(max) $> 20$ mK and $T_B$(max) $> 5$-$\sigma$, where $\sigma$ is
the (locally-defined) noise level at the velocity of each cloud candidate. 
\end{enumerate}

For more details on the cataloging see Putman \etal\ (2002) and de Heij \etal\ (2002).

The spatial distribution of the cataloged HVCs in the vicinity of the
Magellanic Stream is shown in Fig.~\ref{fig9} with crosses representing
positive velocity clouds and triangles representing negative velocity clouds.  
The solid circles represent the galaxies of the Sculptor Group (discussed below).  The
catalog method was originally designed to catalog compact clouds and large
complexes such as the Stream are divided into a number of small clouds.  This
is evident from a comparison of Fig.~\ref{fig9} and Fig.~\ref{fig5}.
Since the catalog method does not include clouds between $\pm 80$ \kms\ and misses some
extended emission due to the spatial size of the cubes used in the cataloging, 
flux is missing from the catalog of Stream clouds.
For instance, in the Stream regions MS~I -- IV (see Fig.~\ref{fig5}), the
total mass of the 364 cataloged clouds is $\sim 1\times 10^7$ \Msun,
which is significantly smaller than the mass of $9.3\times 10^7$ \Msun\ tabulated in
Table~\ref{tbl2}.  The difference is partially due to $4.7\times 10^7$ \Msun\
of the Stream in this region lying at uncataloged velocities (between 
$v_{\rm LSR} = \pm 0$ -- 80 \kms).  Some of the rest of the Stream may not be
in the HVC catalog because the clouds eventually merge with gas between $\pm 80$ \kms.  
Fig.~\ref{fig9} still shows a large number of cataloged clouds along the Stream and a 
particular abundance of clouds around $L_M
\approx$ 310\deg.  There is also a large number of cataloged clouds
between $L_M =$ 0\deg\ to 30\deg, but some of these represent low Galactic latitude 
high-velocity clouds which are not included in the integrated intensity maps of
Fig.4 and \ref{fig5}.

The properties of the cataloged clouds along the Stream are presented
in Fig.~\ref{fig10}, with the clouds between 260\deg\ $< L_M <$ 360\deg\ and -25\deg\ $< B_M <$ +15\deg\ considered Stream clouds.  The
properties of the clouds in the Sculptor region between 295\deg\ $< L_M <$
320\deg\ and -25\deg\ $< B_M <$ +15\deg\ are also shown independently
as the open circles, and are described at the end of this section.  The
lower limit on the peak brightness temperature for the clouds shown is
0.04 K.  For the maximum column density, the lowest value is $N_{HI}
\approx 2\times 10^{18}$ cm$^{-2}$, corresponding to a cloud with a
velocity width of $\sim 30$ \kms\ at the peak brightness temperature
limit.  The lowest flux of the clouds shown is 2.5 Jy \kms, which
corresponds to a mass limit of $1.8 \times 10^3$ \Msun\ at 55 kpc.
The catalog may only be complete at HI fluxes of 25 Jy \kms\ and higher
however (see Putman et al. 2002).
The sizes of the clouds are limited by the 15.5\arcmin\ survey
resolution, and the velocity width limit is set by the 26.4 \kms\
velocity resolution.
Spatially, the cataloged clouds show a slight tendency for a decrease in total \HI\
flux and maximum column density from the head to the tail of the
Stream; however, the majority of the
cloud properties remain approximately constant along the Stream.
The cataloged clouds along the Stream
have typical masses of $1.2 \times 10^4$ \Msun\ and are half a degree
in size, with a median maximum column density of $1.0 \times 10^{19}$
cm$^{-2}$.

For the clouds along the Stream, the differential \HI\ flux
distribution is $f(\log F_{HI}) \propto F_{HI}^{-1.0}$
(Fig.~\ref{fig10}a), or $f(F_{HI}) \propto F_{HI}^{-2.0}$, which is
steep and implies that clouds of all \HI\ masses contribute
significantly to the total \HI\ mass.  
Unlike the entire population of HVCs, the Magellanic HVCs can
probably be placed at somewhat similar distances and the total flux distribution
can be thought of in terms of a mass distribution.  This HI mass
distribution is steeper than that of galaxies (Kilborn et al. 2002; Rosenberg \& Schneider 2002).
The peak column density distribution function, $f(N_{HI}) \propto
N_{HI}^{-2.8}$ (Fig.~\ref{fig10}b), similarly implies the \HI\ resides
in clouds with a wide range of column densities, though dominated more (in
numbers and total HI mass) by
clouds of low $N_{HI}$ than high $N_{HI}$.  This column density distribution is steeper
than that of galaxies (Zwaan, Verheijen \& Briggs 1999) and Ly$\alpha$ absorber systems (Penton et al. 2000).  The solid-angle distribution 
function, $f(\Omega_{HI}) \propto \Omega_{HI}^{-2.0}$, is shown in Fig.~\ref{fig10}c.
The semi-major axis distribution function derived from the 25\% of peak column
density contour is $f(R_{HI}) \propto R_{HI}^{-3.2}$.  
The solid angle distribution implies that coverage is provided by small clouds and large clouds in nearly
equal measure.  The velocity width distribution function is extremely
steep, $f(W_{50}) \propto W_{50}^{-6.3}$ (Fig.~\ref{fig10}d), with the
great bulk of clouds having small velocity widths. This distribution
function is not well-measured in the HIPASS data because of the coarse
velocity resolution.  At higher velocity resolution it is likely that
these velocity widths would generally decrease.  Another selection effect
which 
affects the slope of this plot is that higher velocity width objects are likely
to have lower peak fluxes and may have been undetected in this catalog.

The velocity distribution of the individual clouds along the Stream is
shown in Fig.~\ref{fig12}. In the LSR frame (Fig.~\ref{fig12}a), the gradient 
is roughly 7.2 \kms\ per degree (see also Cohen 1982).
With respect to a Galactic standard of rest (Fig.~\ref{fig12}b) or a
Local Group frame (Fig.~\ref{fig12}c), the gradient along the Stream
is less apparent, indicating that it does not appear to participate in
Galactic rotation.  However, a gradient remains and large sections lie at predominantly negative
velocities, as discussed in the previous section. At $L_M \approx$
310\deg, near the South Galactic Pole (SGP), there is a slight hiccup in the
Stream's velocity gradient which may be due to a population of clouds not related
to the Stream (see below).  Despite the fact that the main filament
of the Stream goes through $v_{\rm LSR} = 0$ \kms\ at this position and
there should be less clouds cataloged, there is a clear overabundance
of clouds at a wide range of velocities.  The spread of cloud velocities in
this region is larger than any other position along the Stream, with clouds 
cataloged between $v_{\rm GSR} = 230$ \kms\ and
$-280$ \kms\ ($v_{\rm LSR} = 250$ \kms\ to $-250$ \kms).  This
includes the most negative GSR velocities along the Stream.  A histogram 
of cloud numbers versus longitude (Fig.~\ref{fig12}d) shows that this region
contains the highest density of clouds anywhere along the Stream, as
is also apparent in Fig.~\ref{fig9}.  
A second region of excess cloud population lies at about $L_M \approx$
10\deg\ (see Fig.~\ref{fig12}d). As Figs~\ref{fig12}a-c show, there
appears to be a bifurcation in velocity at this position, most likely
representing a population of clouds leading the Magellanic system and
a contaminating Galactic population.  The clouds in this region all
lie at positive LSR velocities, peaking at $\sim 370$ \kms\ close to
the LMC.  They also have the greatest range of \HI\ fluxes.

As mentioned above, the region about the South Galactic Pole, or
295\deg\ $< L_M <$ 320\deg, is of special interest.  This region
contains an overabundance of cataloged clouds with a wide range of
velocities, and is also where the Magellanic Stream passes in front of
the galaxies of the nearby Sculptor Group (Fig.~\ref{fig9} and
\ref{fig14}) and through 0 \kms\ (Fig.~\ref{fig15}).  Because the
catalog does not include objects between $\pm$ 80 \kms, clouds which
would be considered canonical members of the Stream are automatically
excluded in an investigation of this region.  The Sculptor group
extends between 295\deg\ $< L_M <$ 320\deg\ and -21\deg\ $< B_M <$
9\deg\ (Cote\etal\ 1997), and in distance between 1.7 Mpc and 4.4 Mpc,
with galaxies at smaller Magellanic longitudes (to the north on the sky) being
more distant.  Fig.~\ref{fig9} shows the distribution of clouds and
Sculptor Group galaxies (from NED) in Magellanic coordinates with the
triangles representing negative velocity clouds, the crosses positive
velocity clouds and the solid circles the Sculptor Group
galaxies which are all at positive velocities.  Note that the overabundance of clouds actually extends
across the Stream, beyond the $B_M$'s of the Sculptor Group galaxies.  
The galaxies in the closer half of the Sculptor
Group ($\sim 2$ Mpc; southern half on the sky) sit exactly in the region of the overabundance of positive and
negative velocity clouds ($L_M \sim 310$\deg).  There does not seem to
be the same overabundance of clouds overlapping with the more distant half
of the group ($\sim2.6 - 4.4$ Mpc), at $L_M \sim 295$\deg.  This is also shown in
Fig.~\ref{fig14}, where positive velocity clouds are scattered amongst
the galaxies in the closer half of the group, while the more distant 
half of the group (closer to the South Galactic Pole) contains only
galaxies.  Fig.~\ref{fig15} shows that Sculptor galaxies have
velocities which vary from 50 \kms\ to 700 \kms, while the clouds have
velocities between $\pm 250$ \kms.  Though these are very different velocity
ranges, a number of
the galaxies do have similar velocities to the clouds.  The Sculptor
dSph also lies in this region (labeled in Fig.~\ref{fig14}), at a
velocity of $\approx$ 100 \kms\ and a distance of 79 kpc (Mateo 1998).

The properties of the clouds in this region of overlap with the
Sculptor Group, represented by the open circles in Fig.~\ref{fig10}, do 
not greatly differ from the properties of the clouds
along the entire Stream.  This makes their HI mass and column density distributions
steeper than that of galaxies and Ly$\alpha$ absorbers, but similar 
to the other HVCs cataloged in the southern sky (Putman et al. 2002).
The primary difference between the two populations of clouds (besides the spatial
and velocity anomalies discussed above) can be found
in the distribution of position angles (Fig.~\ref{fig11}).  The position angles for the elongated clouds
in the region of the Sculptor Group do not show the same bias to be
aligned with long axis of the Stream as the rest of the elongated Stream clouds.  
The distribution of the clouds with well-defined position angles
(minor-to-major axis ratio less than 0.7) is shown in Fig.~\ref{fig11}
in terms of their Magellanic position angle. The top panel shows a broad
peak at pa$_M$
$\sim$90\ deg\ which shows that the elongated clouds with 360\deg\ $> L_M >$ 260\deg\
are largely aligned with their major axes parallel with the Stream.  78 of 
the 270 elongated clouds lie between pa$_M$'s 75\arcdeg and 
105\arcdeg, which is 70\% more than expected from a random distribution. 
On the other hand, only 11 of the 98 elongated clouds in the region of the Sculptor Group
lie in this range of pa$_M$'s, which is close to the number expected from a
random distribution.

\section{Origin of the Stream}

While it remains clear that the Stream has originated from the
Magellanic Clouds, there remain several plausible formation scenarios.  
In the traditional tidal model, the Stream was
pulled from the SMC 1.5 - 2 Gyr ago during a previous perigalactic
passage and a close encounter between the two Clouds.  The Leading Arm
shows that tidal forces play an important role in this interaction,
but it is clearly not as simple as the current models may suggest.
There remain a number of features which the tidal models have yet to
explain. 

 To begin, the column density distribution does not gradually drop off
towards the Stream's tail in the tidal simulations (e.g. Gardiner \&
Noguchi 1996; hereafter GN96).  Though the HIPASS observations show the Stream to be
extremely clumpy, there is still a systematic decrease in the mean
column density from MS I to MS VI (see Table~\ref{tbl2}) which should
be explained.  The tail (MS VI) is the oldest part of the Stream and
the lower column densities could represent the nature of the first
material which was pulled from the Clouds' outer halos or simply the
gas having more time to disperse.  Cloud evaporation due to
interaction with the Galaxy's halo may also be important.   Though ram
pressure forces are unable to form the Stream and Leading Arm using
reasonable halo densities, they may be responsible for
shaping it (e.g. Murali 2000; Li 1999).   A small
amount of ram pressure (using halo densities on the order of $5 \times 10^{-5}$ cm$^{-5}$)
added to the tidal models may also explain the
mass, structure and deflection angle of the Leading Arm (Gardiner 1999; Putman et al. 1998). 

A model which successfully represents the Magellanic Stream should
also address the origin of the numerous small filaments which emanate
from the top of the SMC and Bridge and the break in the Stream at $L_M
\approx 330$\deg.  These features can be seen most readily in
Figs.~\ref{fig4} and ~\ref{fig5}, with the end of the multiple filaments and the break
at approximately the same position.  Since the Clouds are most likely
currently close to perigalacticon, this web of gas at the beginning
of the Stream may represent freshly stripped material from the Clouds.
The complexity of the gas may be due to both the clumpiness of the gas
pulled from the Clouds (see the \HI\
maps of the LMC (Kim \etal\ 1998) and SMC (Stanimirovic \etal\ 1999)), and the binary nature of
the Large and Small Cloud which has not allowed for the formation of a
smooth tidal tail in this region.  A similar past loosening of gas from the
Clouds most likely led to the formation of the Stream's long narrow filaments.
It is possible that the material at the head of the Stream is falling back onto the Clouds, however, although this gas is at low
relative velocities to the Clouds, it follows the same velocity
gradient as the rest of the Stream.

The majority of the Stream is made up of dual filaments.  As discussed in
section 3.1, this may be due to concentrations of gas being pulled
separately from both the SMC and the Bridge; however, there are several other
possibilities.  Mathewson and Ford (1984) were the first
to point out that the SMC appears to be made up of two
velocity-separated concentrations of neutral gas, and they argued that
the SMC was ripped into two entities during a close encounter with the
LMC.  One could imagine each filament representing one of these
entities; however the velocity separation between these components is
40 \kms, and though this kinematic separation is seen in components of
the Magellanic Bridge, it is not seen in the Stream.  The two
filaments are at approximately the same velocity and appear to be 
periodically twisting about each other.  Dual filaments
are reproduced in the GN96 model, both originating from the SMC, but
they are at much greater spatial separation.  GN96 do not offer an
explanation for their presence and the dual filaments are not
reproduced in the tidal simulations of Yoshizawa (1998). 
The double helix structure is
suggestive of some type of magnetically driven structure (e.g. the
Galactic Center filaments; Yusef-Zadeh\etal\ 1997), but this is merely
an observation, as there is unlikely to be a strong field or enough
non-neutral material to produce magnetically driven structures.
The complex 3-body interaction between the Clouds and the Galaxy is most likely
responsible for the Stream's double helix structure, as the Clouds are rotating
about each other as they rotate about the Galaxy.  If the dual filaments
have originated from both the Bridge and SMC, it implies that the Bridge
(or something similar to the currently observed Bridge) is
actually older than the Stream (rather than younger as suggested by many models, $\sim 0.2 $ Gyr) and that the Clouds have been bound together for
at least one or two orbits about the Galaxy.

Under certain circumstances (i.e. using IGM densities as high as a 
few $\times 10^{-2}$ cm$^{-3}$) ram pressure
stripping can produce dual filaments at the same velocity.  This
happens when the outer material is swept behind the galaxy as it moves
through a viscous medium (e.g. Quilis, Moore \& Bower 2000).  The ram
pressure simulations show that the two filaments eventually connect at
some locations, much like the periodic connections between the two
filaments seen here.  There are several problems with this model.  The survival
timescales of the gas with this high of halo density is too short to create the
100\deg\ long Stream, and since material is swept from all sides of
the moving galaxies, lower column density material would also be
present in the center of the two filaments.  This is not observed at
the current detection level.  Also, though ram pressure forces can cause
material to decelerate, fall in its orbit and lead the clouds, it is
unlikely to leave the continuous Leading Arm of gas partially shown in 
Fig. 4.   Finally, another mechanism which could produce the features
of the Stream but requires very high halo densities is a wake process as
the Magellanic Clouds move through a hot Galactic halo (Liu 1992).  

 An important new characteristic of the Stream is the small
semi-isolated clumps of \HI\ which surround the main filaments in position
and velocity.  They
are especially ubiquitous at the head of the Stream and along the
sightline discussed in the next section.  These clouds may represent
the clumpiness of the original gas, or instabilities along the
Stream's edges as it struggles to flow through a diffuse medium in the
Galaxy's halo (Pietz~\etal\ 1996).  Some of the gas concentrations on
the leading side of the Magellanic Clouds may have been scattered off
the Leading Arm and accelerated to high velocities by the Galaxy
and LMC, as predicted by the hydrodynamical models of Li (1999).  The
head-tail structure of many of the clouds indicates they are either
evaporating or being continually tidally stretched.  The tails do not
always point in a single direction, but as shown in Fig~\ref{fig11},
they are preferentially elongated along the Magellanic Stream,
indicating that most of these clouds are related to the
Magellanic System.
If one assumes that the clumps were pulled out with the Stream and are
1-2 Gyr old, the length of the tails (which typically contain half the
mass of the central clump) could be used to estimate the density of
the Galaxy's halo or the tidal field history.  Quilis \& Moore (2001) find
head-tail features are created in HVCs with typical lifetimes of $\sim$ 1
Gyr as long as the Galactic wind density
is higher than $10^{-4}$ cm$^{-3}$.  They predict that at the distance of
the Magellanic Clouds and beyond, the wind density will have decreased to
$< 5 \times 10^{-5}$ cm$^{-3}$ and head-tail structures would not be
created if the clouds are dark matter dominated.  They also find a much
higher head to tail mass ratio ($\sim$ 10:1) than observed in the pure gas
cloud models.
The presence of large
head-tail features which appear to be associated with the Magellanic Stream
argues that the halo density at 50 kpc may be higher than expected and/or 
the clouds are diffuse structures which do not contain a significant amount of
dark matter.

The head-tail features extend up to $\sim$ 20\deg\ from the main filament of
the Stream and they could also be
cold concentrations within an extended, mainly ionized Stream component.  This is
supported by sensitive HI emission line observations (Lockman et al. 2002) and Mg II
absorption line observations (Gibson et al. 2000) which find low
column density extensions more than 10\deg\ from the HI limits
of the tail of the Stream shown here.  \Ha\ observations along the Stream
have detected variable degrees of emission from the head to the tail (e.g. Weiner \& Williams 1996; Putman 2000; Madsen pers. comm.).
Explanations for the origin of this emission remain unclear, but future
\Ha\ emission and UV absorption line observations should clarify the extent
and ionizing mechanism of the Stream.  

A final issue for the Stream's tidal origin is the lack of corresponding stellar
tidal debris (e.g. Moore \& Davis 1994; Guhathakurta \& Reitzel 1998).
Are there stars associated with the Stream which have been missed
due to the limitations of previous surveys?
Despite the insistence that stars should have been stripped off with
the gas in the tidal models, the observations show that stellar tails
are often completely missing or offset from the gaseous tails in
interacting systems (e.g. the M82 group, Yun\etal\ 1994; Hibbard, Vacca \& Young
2000) and this can be reproduced in simulations (Barnes 2002).   This is
especially true when the galaxy's initial gas distribution is more
extended than the stellar component (Mihos 2001; Yoshizawa 1998; Li 1999), a common situation
(Broeils \& van Woerden 1994).  The possibility of an offset stellar stream made up of old
stars also still exists (Majewski \etal\ 1999; van der Marel 2001).

\section{Origin of the Clouds along the Sightline to the Sculptor Group}

The \HI\ observations presented here show that the clouds in the region of
the South Galactic Pole, or Magellanic longitude $\sim$310\deg, are 
unlike the clouds along the rest of the Stream in several ways.   Their velocity distribution does
not match the Stream's velocity gradient and the number of clouds does
not agree with the normal abundance of clouds found along the Stream.
This group of clouds happens to lie in the same region as the southern (on the sky)
and near half of the Sculptor Group.  Mathewson\etal\ (1975) were the
first to point out the coincidence between these anomalous velocity
clouds and the Sculptor Group galaxies.  They argued that the
clouds towards the galaxies NGC 55 and 300 were associated with the
Sculptor Group rather than the Magellanic Stream based on the kinematic
and positional similarities.  This origin was countered by Haynes \&
Roberts (1979) and Haynes (1979), who argued several points against
the clouds being members of the Group, but did not provide an
explanation for the anomalous velocities of the gas.  Here, we
re-address several of the points made by Haynes\etal, as well as
introduce additional observational constraints and investigate possible origins.  This excess population
of clouds in the region of the near half of the Sculptor Group will
be referred to as the anomalous clouds in this discussion.

\subsection{Relationship to the Sculptor Group Galaxies?}

An important constraint on the anomalous clouds having a relationship 
to the Sculptor Group is their continuation 
beyond both the spatial and velocity limits of the
galaxies in the group (see Figs.~\ref{fig9} and \ref{fig15}).  
As shown in Fig.~\ref{fig9} these clouds do not solely overlap with the Stream 
and near half of the Sculptor Group (galaxies are the solid points), but form a broken string of
material which crosses the Stream almost perpendicularly and extends
at least 25\deg~ from the Stream.  The Sculptor Group galaxies cluster about 
Magellanic latitude $\approx
-5$\deg, while the anomalous clouds actually continue in a string to the limits 
of Fig.~\ref{fig9}.  In velocity space (Fig.~\ref{fig15}), the anomalous clouds range from approximately
-250 \kms\ to +250 \kms (LSR), while the Sculptor Group galaxies extend from
50 to 700 \kms, with no galaxies at negative
velocities.  The anomalous clouds are not solely small and
compact as might be expected if they were at large distances, but show diffuse connections between each other and a
signature of a velocity gradient as they extend across the Stream
(see Fig.~\ref{fig7}).  Fig.~\ref{fig10} also
shows that the flux, column density, size and velocity width properties of these clouds do not greatly
differ from the bulk of the Stream clouds.  These results
suggest that the anomalous clouds are a chance superposition onto the 
near half of the Sculptor Group which was picked
out as interesting due to the limited area and sensitivity covered by previous
surveys. 

A further constraint on a relationship between the anomalous clouds and the
Sculptor Group galaxies is the lack of cloud detection in the more distant half
of the group ($L_{M} \approx 295$\deg).  This was also
noted by Haynes (1979), but it was unclear if this was a significant
discriminant considering the further distance.  The lack of clouds in this 
region of the Sculptor group is most apparent in the integrated intensity 
map of Fig.~\ref{fig14} about the galaxies at the South Galactic Pole.  
The cloud-free half of the group ranges in distance from 2.6 Mpc
(e.g. NGC 247 and NGC 253) to 4.4 Mpc (e.g. NGC 45 and NGC 59).
HIPASS is sensitive to HVCs with masses above $\sim 0.36$ D$^2$ (kpc) \Msun\
($5\sigma$; $\Delta V=$35~\kms), or $> 2.4 \times 10^6$ \Msun\ at 2.6
Mpc and $> 7.0 \times 10^6$ \Msun\ at 4.4
Mpc.  If the clouds towards the closer half of the Sculptor Group ($L_M \approx$
310\deg) were at
the distance of those Sculptor Group galaxies ($\sim$2 Mpc) their HI
masses would range from 2.8 - 200 $\times 10^6$
\Msun.  Therefore if intergalactic \HI\ clouds are
present in the more distant half of the group, most of them should have been
detected, despite the greater distance.  The cloud distributions shown in Fig~\ref{fig9} 
and \ref{fig15} may be affected by the completeness limitations of the cloud catalog.
However, at 2.6 Mpc the catalog should be complete to HI masses of $4 \times\ 10^7$ \Msun, and
so if a similar population of clouds was in the more distant half of the Sculptor
Group, almost half of them should have been cataloged.

It is possible that leftover intergalactic \HI\ clouds would not be
uniformly distributed throughout a group due to the distribution  of
the galaxies.  For instance, in simulations of the formation of the
Local Group, the presence of M31 and the Milky Way leads to a long
filamentary bias in the distribution of leftover satellites (Moore et al. 2001).  
This argument does not justify the asymmetric
distribution of clouds presented here considering the massive
galaxies in the northern half of the Sculptor Group.  However, since
the Local Group and the Sculptor Group appear to form a filament of
galaxies, rather than two unique groups (e.g. Jerjen, Bingelli \&
Freeman 1998), the bias in the clouds distribution towards the
southern and closest part of the group could actually be an extension
of a population of Local Group intergalactic \HI\ clouds in the
direction of the closest group to the Local.  This explanation places
the detected clouds closer and makes them less massive (generally $< 6
\times 10^6$ \Msun\ at 1 Mpc),  leaving the non-detection in the
northern half of the Sculptor Group and other groups open to further
speculation (e.g. de Blok et al. 2002; Zwaan 2001; Banks\etal\ 1999).
Though the diffuse connections between the clouds was used to argue
for a closer origin above, extended \HI\ at lower column densities
might also be expected if the clouds are the equivalent of the
Ly$\alpha$ absorbers along filaments of galaxies.  On the other hand,
explaining the kinematic and spatial extension of the clouds beyond
the Sculptor Group is not straightforward.  

\subsection{Alternative Explanations}

A possible origin for the anomalous clouds
which may be related to their position about the South Galactic Pole, rather
than their coincidence with the Sculptor Group, is the passage of previous
Galactic satellites through the Galactic Poles which has produced
crossing tidal streams.  Though the anomalous string of clouds is very
different from the Magellanic Stream, the velocity gradient
of the clouds as they extend across the Stream supports this type of 
scenario (Fig.~\ref{fig7}).   
Satellites to spiral galaxies appear to have a preference for polar orbits.
This was first stated by Holmberg (1969) 
and has been confirmed for the Milky Way (e.g. Lynden-Bell 1982;
Majewski 1994) and for other Milky Way-like spirals (Zaritsky\etal\ 1997).
Zaritsky \& Gonzalez (1999) looked for a dynamical process which could be
responsible for this anisotropy through galaxy formation models which
start with a spherically symmetric distribution of matter.  They found
that the would-be satellites are preferentially removed  from
low-inclination orbits, leading to the current preference for a polar
distribution of satellites.  

Polar orbits for the Milky Way satellites should lead to a 
preferential distribution of tidal debris about the Galactic Poles.
Crossing tidal streams are a common prediction which is highly
likely in our Galaxy's complex halo (e.g. Lynden-Bell \& Lynden-Bell
1995).   A candidate satellite which could be responsible for the trail of \HI\ clouds
presented here is the Sagittarius dwarf galaxy.
The Sgr dSph currently lies $\approx$ 16 kpc from the Galactic Center
(Ibata, Gilmore \& Irwin 1994) and is generally believed to have a
polar orbit like the Magellanic Clouds, but nearly orthogonal
to the Clouds orbit (Ibata\etal\ 1997).  The period of the dwarf's
orbit is $<$ 1 Gyr and its extended structure suggests that it is
currently being torn apart by the tidal forces of the Milky Way.
Though the Sgr dSph does not currently contain neutral hydrogen
(Koribalski\etal\ 1994),  it surely did within the past few Gyr
considering that star formation was occurring within 0.5 to 3 Gyr ago
(Layden \& Sarajedini 2000).   If the dwarf once had extended
neutral hydrogen gas, it is likely to have been the first thing
stripped from it and any remnant would be at large radii from the
dwarf's current position.  Future work will need to look at the full
extension of the clouds about the Galaxy and make direct comparisons
to models of the Sgr dSph orbit (e.g. Ibata \& Lewis 1998;  Helmi \& White 2001).

As shown in Fig.~\ref{fig14}, the Sculptor dSph also lies in a spatial
and velocity position which overlaps with the Stream and may have some
relationship to some of the clouds (e.g. Carignan 1999).  It is at a
distance of 79 kpc and if tidal models are correct in their prediction
that the Stream becomes more distant as it moves northward, the Scl
dSph could  be within 10 kpc of currently being embedded in the Stream.
Schweitzer (1996) found that the Scl dSph has a space velocity of 220
+/- 125 \kms\ roughly towards the Fornax dSph, or almost perpendicular
to the  Magellanic Stream.    Irwin \& Hatzidimitriou (1995) also
suggest that the Scl dSph is undergoing tidal disruption based on the
stellar data.   Another possibility is that the Scl dSph has actually
formed out of the Stream's debris, but the old age of its stellar
population (Grebel 1996) and proper motion indicate this is unlikely.

Finally two other possibilities for the clouds' origin should be mentioned.
The first is a possible direct link to our Galaxy through some type of
Galactic fountain flow.  Reproducing the velocities of the clouds
would be difficult with this model (e.g. Wakker \& Bregman 1990), and placing the clouds
over the Galactic Center requires the remnants to
be at the distance of the Magellanic Stream.
If the clouds are directly related to the 
Galaxy, near solar metallicities and distances less
than 10 kpc would be expected.  \Ha\ emission may also be detected (Bland-Hawthorn \etal 1998).
The second possible origin for the anomalous clouds is a direct relation to
the interaction of the Magellanic Clouds with themselves and the Milky Way.
The beginning of the Stream is very chaotic, most likely representing the fresh stripping of material as the Clouds pass through perigalacticon.  The
multiple filaments at the head of the Stream may have interacted and resulted in a broad
velocity range of debris which has extended over a larger region of
sky than most of the Stream.  
The difference in the distribution of position angles
(Fig.~\ref{fig11}) would also have to be explained in this scenario.
Leading Arm material which has wrapped all the way around to overlap
with the Stream could be responsible, but a direct connection between the
anomalous clouds and the Leading Arm
has not been found in the HIPASS data.  Finally the anomalous Stream clouds 
could represent what is left from a previous passage of the Magellanic Clouds through the South Galactic Pole.

\section{Overview}

The Milky Way contains between 2 - 4 $\times 10^9$ \Msun\ of neutral
hydrogen (Sparke \& Gallagher 2000) and when the \HI\ masses of the
LMC, SMC, Bridge, Stream and Leading Arm are combined (masses of 4.8
$\times 10^8$ \Msun; Staveley-Smith \etal\ 2002,  4.2 $\times 10^8$ \Msun;
Stanimirovic \etal\ 1998, $5.5 \times 10^7$ \Msun, $2.1\times 10^8$
\Msun; Putman \etal\ 1998, and $2 \times 10^7$ \Msun\ respectively)
they make up one third to one half the \HI\ mass of the Milky Way.  The
interaction between the Galaxy and the Magellanic Clouds has created a
variety of unique \HI\ features which can be viewed in detail due to its
proximity.  The Stream is bifurcated along its length and this has most
likely originated from separate filaments being pulled from both the SMC
and Bridge.  This 
suggests the Bridge (or something similar to the currently observed Bridge)
is older than the Stream and also older than conventionally assumed.  Dense HI
clouds, which match the spatial and velocity gradient of the Stream, surround
the main filament and
indicate that debris from the interaction extends out to 20\deg\
from the Stream's main filament.  The dense clumps of Stream debris
commonly have tails which suggest they are
undergoing an interaction with the Galaxy's halo.  The length of the tails
indicates these clouds are not dark matter dominated or the halo
densities at the distance of the Magellanic Stream are higher than
$10^{-5}$ cm$^{-3}$.  A small amount of ram pressure stripping 
may also explain the column density drop off along the Stream.
Many of the extreme negative and positive high-velocity
clouds have been associated with the Magellanic System in this paper and
this identification permits 
Galactic origins for the remaining population of HVCs, since these models
cannot reproduce many of the highest HVC
velocities.  

The abundance and velocity range of the high-velocity clouds along the
sightline to the less distant half (southern half on the sky) of the Sculptor Group is not seen in
other parts of the sky (Putman et al. 2002; WvW97) and may be due to the position of the
clouds at the South Galactic Pole.  It has been shown that satellites prefer polar orbits about
spiral galaxies, and this should result in a preferred distribution of
satellite debris about the Galactic Poles.  It is unlikely that the clouds are
members of the Sculptor Group considering the velocity and spatial
distributions presented here and the lack of
detection of similar clouds in the northern half of the group.  This result argues
against a Local Group infall model for HVCs (e.g. Blitz\etal\  1999).
The Sculptor Group has commonly been referred to as the nearby Local
Group equivalent which is also unvirialized and possibly falling
together for the first time.  In this respect, if HVCs are scattered
throughout the Local Group and represent the leftover infalling
building blocks, the building blocks of the Sculptor Group galaxies
should have also been found.  The origin of the anomalous clouds can be
clarified with distance and metallicity determinations.

\acknowledgements{We are grateful to Mark Calabretta, Malte Maquarding, Neil Killeen and the aips++ crew at ATNF for
their help with the data reduction and imaging.  We also thank Vincent de Heij for
creating the final mosaic.  The Multibeam
Working Group is acknowledged for their assistance with the
HIPASS observations.}

\clearpage


\clearpage

\begin{figure}

\caption{Channel image ($v_{lsr}$ $\approx$ 220\kms) of a section of the Magellanic Stream reduced with 
the {\sc minmed5} method (top) and the standard HIPASS method (bottom). Negative
 (grey) and positive (black) contours 
are shown for $\pm$ 0.05, 0.1, 0.2, 0.4 and 0.8 K.}
\label{fig:hvcchan}

\end{figure}

\begin{figure}
\caption{
Integrated intensity image of a section of the Magellanic
Stream which shows the results of the HVC {\sc minmed5} reduction (top)
and the flux filter applied with the standard HIPASS reduction (bottom).
This is the same region of sky as
Fig.~\ref{fig:hvcchan}.  The column density contours
are 0.5, 1, 2, 4, 8 and 16 $\times 10^{19}$ cm$^{-2}$.  Values less than
0.01 K have been excluded from this map.}
\label{fig:hvcmom}
\end{figure}

\begin{figure}

\caption{The spectrum along a single sightline in the Stream (towards Fairall 9) to
show the distinct difference in the {\sc minmed5} reduced spectrum (solid line) and
the standard HIPASS reduction spectrum (dashed line).}
\label{fig:fair9comp}
\end{figure}

\begin{figure}
\caption{Integrated intensity map over 8000 deg$^2$ of southern sky showing the Large Magellanic Cloud (LMC), Small Magellanic Cloud (SMC), Magellanic Bridge, Magellanic Stream and part of the Leading Arm as seen with the {\sc minmed5} reduced HIPASS data.}
\label{fig:stream}
\end{figure}

\begin{figure}
\caption{Annotated \HI\ column density 
image of the Magellanic Clouds, Bridge,
Stream and the beginning of the Leading Arm feature (LAF).  Velocities from
-450 to 400 \kms\ are included, excluding $\pm$20\kms\ due to confusion 
with Galactic emission.   The boxes represent the regions used for
the mass determinations in Table~\ref{tbl2}.  Magellanic longitudes for positions
along the Stream are also labeled.
The intensity values are on a logarithmic scale with black corresponding to
$N_{HI} > 6 \times 10^{20}$\,cm$^{-2}$ and the faintest levels 
corresponding to $\sim 2 \times 10^{18}$\,cm$^{-2}$.}
\label{fig5}
\end{figure}

\begin{figure}
\caption{Channel maps of HI emission over
8000 deg$^2$ of the southern sky showing the Magellanic System and the
Galaxy. The tail of the Stream is at $\ell=90\arcdeg$, $b=-50\arcdeg$, $v_{\rm LSR}$
$=-343$ km s$^{-1}$, and the head of the Stream (where it leaves
the Magellanic Clouds) is at  $\ell=295\arcdeg$, $b=-45\arcdeg$,
$v_{\rm LSR}$ $=185$ km s$^{-1}$. Channel maps are shown every 26.4 km s$^{-1}$.
The LSR velocity is displayed at the top-left of each panel. At velocities
between $-53$ and 53 km s$^{-1}$, Galactic emission is prominent but is
not accurately represented because of the observation and data reduction
method. However, the main Stream filament is still apparent at these
velocities. Only partial HIPASS data was utilised for the northern
declination bands (Decl. 2\arcdeg\ to 25\arcdeg), resulting in the elevated
noise levels apparent at the right hand side of each panel. Only an example channel is in the astro-ph version. }
\label{fig6}
\end{figure}

\clearpage
\begin{figure}
\caption{Velocity distribution of the Magellanic System ranging from
$v_{\rm LSR}$ = $-$450\,\kms\ (light grey) to $+$380\,\kms\ (dark). }
\label{fig7}
\end{figure}

\begin{figure}
\caption{Integrated intensity maps of head-tail clouds found along the Stream.  
Contours begin at 2 $\times 10^{18}$ cm$^{-2}$ and increase in increments of 3 $\times 10^{18}$.
The arrow points along the direction of the long axis of the Magellanic Stream.} 
\label{fig8}
\end{figure}

\begin{figure}
\caption{Spatial distribution of the cataloged clouds along
and around the Stream 
and the galaxies of the Sculptor Group (solid points) in Magellanic coordinates.  
The positive velocity clouds are crosses and the negative 
velocity clouds are triangles.} 
\label{fig9}
\end{figure}

\begin{figure}
\caption{Properties of the 486 cataloged HVCs 
about the Magellanic Stream (-25\deg $< B_M < +15$\deg, and
260\deg\ $<$ $L_M$ $<$ 360\deg; solid points): (a) \HI\ flux; 
(b) peak column density;
(c) solid angle; and (d) velocity width of the central spectral profile.
Power law fits (solid lines) are overlayed. 
The HVCs are binned logarithmically, so a power law fit 
$f(\log X)\propto X^{\gamma}$ implies $f(X)\propto X^{\gamma-1}$ (see text).
Distributions for clouds along the Sculptor Group sightline (295\deg\ $< L_M < 320$\deg)
are shown by the open circles. }
\label{fig10}
\end{figure}

\begin{figure}
\caption{Magellanic longitude against (a) $v_{\rm LSR}$; (b) $v_{\rm GSR}$;
(c) $v_{\rm LGSR}$; and (d) the number of clouds in each Magellanic longitude
bin for clouds between -25\deg\ $< B_M <$ +15\deg in Fig.9.  Because of confusion
with Galactic gas, no clouds were cataloged between $|v_{\rm LSR}| = \pm80$\kms. 
This velocity limit is overlayed as a light grey line in (a)--(c).}
\label{fig12}
\end{figure}

\begin{figure}
\caption{Integrated \HI\ intensity map of a sub-region of the Stream,
covering $+110 < v_{\rm LSR} < +550$ \kms.
Several galaxies of the Sculptor
Group are labeled, as well as the position of the Sculptor dSph.
The bifurcation of the Magellanic Stream is apparent, as is the abundance
of positive velocity clouds scattered amongst the less distant (southern
on the sky; NGC55, NGC300) galaxies of the Sculptor Group.  
The intensity scale is logarithmic with black regions corresponding
to $N_{HI} > 6 \times 10^{20}$\,cm$^{-2}$.}
\label{fig14}
\end{figure}

\begin{figure}
\caption{Velocity distribution of the cataloged clouds along the Stream (-25\deg\ $< B_M <$ +15\deg; triangles) and the 
Sculptor Group galaxies (solid circles).  Most Sculptor Group galaxies have velocities significantly
larger than any of the cataloged clouds.}
\label{fig15}
\end{figure}

\begin{figure}
\caption{The top plot shows the histogram of Magellanic position angles 
of the 270 elongated clouds (minor-to-major axis ratios $<$ 0.7) along the Magellanic Stream (260\deg\ $<$ $L_M$ $<$ 360\deg\
and -25\deg\ $<$ $B_M$ $<$ +15\deg).
pa$_M=0$\deg\ is perpendicular to the Stream and pa$_M=90$\deg\ is parallel to the Stream.    The bottom plot shows the more uniform
distribution of the 98 elongated clouds 
in the region of the Sculptor Group (295\deg\ $<$ $L_M$ $<$ 320\deg).}
\label{fig11}
\end{figure}

\clearpage

\begin{deluxetable}{lc}
\tablecolumns{2}
\tablecaption{Survey Parameters
\label{tbl1}}
\tablewidth{4.0in}  
\tablehead{
\colhead{Parameter} & \colhead{Value}
}
\startdata
Telescope & Parkes 64\,m   \\
Receiver  & 21-cm multibeam \\
Declination Range\tablenotemark{a}, $\delta$ & $-90\arcdeg$ to $+25\arcdeg$\\
Velocity Range\tablenotemark{b}, $v_{\rm LSR}$ & $-700$ to 1000 \kms \\
Spatial Resolution\tablenotemark{c} & 15\farcm5  \\
Velocity Resolution\tablenotemark{d} & 26.4 \kms  \\
Sensitivity\tablenotemark{e}, $\Delta T_{B}$  ($5\sigma$) &  0.035 K  \\
\HI\ column density limit\tablenotemark{f}~ ($5\sigma$) & $2.2 \times 10^{18}$ cm$^{-2}$  \\
\HI\ mass limit\tablenotemark{g}~ ($5\sigma$) & $1.1 \times 10^3\ (d/55\,{\rm kpc})^2$ \Msun  \\

\enddata
\tablenotetext{a}{Southern HIPASS covers $\delta < 2\arcdeg$ and the
northern extension covers $2\arcdeg \le \delta < 25\arcdeg$. The northern
extension data used here has $\sim 20$\% of the final HIPASS integration
time, and is therefore of lower sensitivity.}
\tablenotetext{b}{Only part of the full HIPASS velocity range, $-1200 < cz < 12700$\,\kms\ is reprocessed.}
\tablenotetext{c}{Varies by $\pm 1\arcmin$ depending on S/N and source size (Barnes et~al. 2001).} 
\tablenotetext{d}{After Hanning smoothing.}
\tablenotetext{e}{Southern HIPASS data.}
\tablenotetext{f}{For an extended cloud with a linewidth of 35\,\kms.}
\tablenotetext{g}{For a point source with a linewidth of 35\,\kms.}
\end{deluxetable}

\clearpage

\begin{deluxetable}{lcc}
\tabletypesize{\scriptsize}
\tablecolumns{3}
\tablecaption{Masses and Mean Column Densities for the Regions 
Defined in Figure~5
\label{tbl2}}
\tablewidth{3.2in}
\tablehead{
\colhead{Object} & 
\colhead{Mass} &
\colhead{$<N_{HI}>$}\\
\colhead{} &
\colhead{[M$_{\sun}$]} &
\colhead{[cm$^{-2}$]}
}
\startdata
\\
\multicolumn{3}{c}{\sc Clouds}\\
\\
LMC\tablenotemark{a}  & $2.9\times 10^8$ & $3.5\times 10^{20}$\\
SMC\tablenotemark{b}  & $3.4\times 10^8$ & $6.5\times 10^{20}$\\
Bridge\tablenotemark{c} & $5.5\times 10^7$ & $1.1\times 10^{20}$\\
Total/mean in region & $6.9\times 10^8$ &  \\
\\
\multicolumn{3}{c}{\sc Stream\tablenotemark{c}}\\
\\

Head of Stream & $1.1\times 10^8$ & $3.3\times 10^{19}$ \\

MS I\tablenotemark{d} & $4.0\times 10^7$ & 
  $3.6\times 10^{19}$\\

MS II\tablenotemark{e}  &   $3.2 \times 10^7$ & 
$3.3\times 10^{19}$\\
 
MS III & $1.4 \times 10^7$ & $2.5 \times 10^{19}$ \\ 
MS IV & $6.7\times 10^6$ & $1.1 \times 10^{19}$ \\ 
MS V &  $3.9\times 10^6$ & $6.2 \times 10^{18}$ \\ 
MS VI &  $1.3\times 10^6$ & $ 3.6 \times 10^{18}$ \\ 
Total/mean in Stream & $2.1\times 10^8$ & $ 2.6 \times 10^{19}$ \\
\\
Total/mean in Clouds and Stream & $9.0 \times 10^8$ & $5.4 \times 10^{19}$ \\
\enddata

\tablenotetext{a}{ LMC assumed to be at 50 kpc.}
\tablenotetext{b}{ SMC assumed to be at 60 kpc.}
\tablenotetext{c}{ Bridge and Stream assumed to be at 55 kpc.}
\tablenotetext{d}{ Mass and mean column density, excluding emission at
Galactic velocities $|v_{\rm LSR}|<20$ \kms, is $3.7\times 10^7$ \Msun\ and
$3.4\times 10^{19}$ cm$^{-2}$, respectively.}
\tablenotetext{e}{ Mass and mean column density, excluding emission at
Galactic velocities $|v_{\rm LSR}|<20$ \kms, is $2.0\times 10^7$ \Msun\ and
$2.2\times 10^{19}$ cm$^{-2}$, respectively. 
Values exclude flux from Sculptor 
group galaxies.}
\end{deluxetable}



\begin{references}
\reference{} Bajaja, E., et al. 1985, ApJS, 58, 143
\reference{} Banks et al. 1999, \apj, 524, 612
\reference{} Barnes, D.G., Staveley-Smith, L., de Blok, W.J.G., et al. 2001, MNRAS, 322, 486
\reference{} Barnes, J., 2002, MNRAS, 333, 481
\reference{} Barnes, J.  \& Hernquist, L. 1996, ApJ, 471, 115
\reference{} Bland-Hawthorn, J., Veilleux, S., Cecil, G.N., Putman, M.E.,
  Gibson, B.K. \& Maloney, P.R. 1998, MNRAS, 299, 611 
\reference{} Blitz, L., Spergel, D.N., Teuben, P.J., Hartmann, D. \& Burton,
  W.B. 1999, \apj, 514, 818 
\reference{} Braun, R. \& Burton, W.B. 1999, A\&A, 341, 437
\reference{} Broeils, A.H. \& van Woerden, H. 1994, A\&A Suppl., 107, 129
\reference{} Br\"{u}ens, C., Kerp, J., \& Staveley-Smith, L. 2000, in Mapping
  the Hidden Universe, ASP Conf. Series 218, eds. Kraan-Korteweg, R.C.,
  Henning, P.A. \& Andernach, H., 349
\reference{} Burton, W.B. 1988, in Galactic \& Extragalactic Radio Astronomy, eds. Verschuur, G.L. \& Kellermann, K.I., Springer-Verlag, 295
\reference{} Carignan, C. 1999, PASA, 16, 18
\reference{} Cohen, R.J. 1982, MNRAS, 199, 281
\reference{} Cote, S., Freeman, K.C., Carignan, C. \& Quinn, P.J. 1997, \aj, 114, 1313
\reference{} de Blok, W.J.G., Zwaan, M.A., Dijkstra, M., Briggs, F.H. \&
Freeman, K.C. 2002, A\&A, 382, 43
\reference{} de Heij, V., Braun, R. \& Burton, W.B. 2002, A\&A, in press
\reference{} Gardiner, L.T. 1999, in Stromlo Workshop on High-Velocity Clouds, ASP Conf. Series 166, 292
\reference{} Gardiner, L.T. \& Noguchi, M. 1996, MNRAS, 278, 191
\reference{} Gardiner, L.T., Sawa, T. \& Fujimoto, M. 1994, MNRAS, 266, 567
\reference{} Gibson, B.K., Giroux, M.L., Penton, S.V., Putman, M.E., Stocke, J.T. \& Shull, J.M. 2000, \aj, 120, 1830
\reference{} Grebel, E. K. 1996, \pasp, 108, 1141
\reference{} Guhathakurta, P. \& Reitzel, D.B. 1998, in Galactic Halos: A UC Santa Cruz Workshop, ASP Conf Series 136, ed. Zaritsky, D., 22
\reference{} Hartmann, D. \& Burton, W.B. 1997, Atlas of Galactic Neutral
Hydrogen, Cambridge Univ. Press
\reference{} Haynes, M.P. 1979, AJ, 84, 1173
\reference{} Haynes, M.P. \& Roberts, M.S. 1979, ApJ, 227, 767 
\reference{} Helmi, A. \& White, S. 2001, \mnras, 323, 529
\reference{} Hibbard, J.E., Vacca, W.D. \& Yun, M.S. 2000, \aj, 119, 1130
\reference{} Hibbard, J.E. \& Yun, M.S. 1999, ApJ, 522, 93
\reference{} Holmberg, E. 1969, Ark. Astron., 5, 305
\reference{} Hu, E. M., Kim, T.-S., Cowie, L. L., Songaila, A., \&
 Rauch, M. 1995, \aj, 110, 1526
\reference{} Hulsbosch, A.N.M. \& Wakker, B.P. 1988, A\&AS, 75, 191
\reference{} Ibata, R.A., Gilmore, G. \& Irwin, M.J. 1994, \nat, 370, 194
\reference{} Ibata, R.A. \& Lewis, G.F. 1998, \apj, 500, 575
\reference{} Ibata, R.A., Wyse, R.F.G., Gilmore, G., Irwin, M.J. \& Suntzeff,
  N.B. 1997, \aj, 113, 634
\reference{} Irwin, M.J. \& Hatzidimitriou, D. 1995, \mnras, 277, 1354
\reference{} Jerjen, H., Freeman, K. \& Binggeli, B. 1998, AJ, 116, 2873
\reference{} Kilborn, V. et al. 2002, AJ, in preparation
\reference{} Kim, S., Staveley-Smith, L., Sault, R.J., Kesteven, M.J.,
McConnell, D., Dopita, M.A. \& Bessell, M. 1998, ApJ, 503, 674
\reference{} Koribalski, B., Johnston, S. \& Otrupcek, R. 1994, \mnras, 270, 43
\reference{} Layden, A.C. \& Sarajedini 2000, ApJ, 119, 1760
\reference{} Li, P.S. 1999, PhD, University of Wyoming
\reference{} Lin, D.N.C., Jones, B.F. \& Klemola, A.R. 1995, ApJ, 439, 
652
\reference{} Liu, Y. 1992, A\&A, 257, 505
\reference{} Lockman, F.J., Murphy, E.M., Petty-Powell, S. \& Urick,
V.J. 2002, ApJS, 140, 331
\reference{} Luks, T. \& Rohlfs, K. 1992, \aap, 263, 41
\reference{} Lynden-Bell, D. 1982, The Observatory, 102, 202
\reference{} Lynden-Bell, D. \& Lynden-Bell, R.M. 1995, MNRAS, 275, 429
\reference{} Majewski, S.R. 1994, \apj, 431, 17
\reference{}Majewski, S.R., Ostheimer, J.C., Patterson, R.J., Kunkel, W.E.,
Johnston, K.V. \& Geisler, D. 2000, AJ, 119, 760
\reference{} Mathewson, D.S, Cleary, J.D. \& Murray, M.N. 1974, ApJ, 190,
291
\reference{} Mathewson, D.S, Cleary, J.D. \& Murray, M.N. 1975, ApJ, 195, L97
\reference{} Mathewson, D.S. \& Ford, V.L. 1984, in Structure \& 
Evolution
of the Magellanic Clouds, IAU Symp., 108, 125
\reference{} Mateo, M.L. 1998, \araa, 36, 435
\reference{} Mebold, U., Herbstmeier, U., Kalberla, P.M.W., Griesen, E.W.,
Wilson, W. \& Haynes, R.F. 1991, A\&A, 251, L1 
\reference{} Mihos, C.J. 2001, ApJ, 550, 94
\reference{} Moore, B., \& Davis, M. 1994, MNRAS, 270, 209
\reference{}Moore, B., Calcaneo-Roldan, C., Stadel, J., Quinn, T., Lake, G., Ghigna, S., \& Governato, F.
 2002, Phys. Rev. D. in press, (astro-ph/0106271)
\reference{} Morras, R. et al. 2000, A\&AS, 142, 25
\reference{} Morras, R. 1985, AJ, 90, 180
\reference{} Morras, R. 1983, AJ, 88, 62
\reference{} Morrison, H.L., Mateo, M., Olszewski, E.W., Harding, P., Dohm-Palmer, R.C.,
Freeman, K.C., Norris, J.E. \& Morita, M. 2000, AJ, 119, 2254
\reference{} Murali, C. 2000, ApJ, 529, L81
\reference{} Newberg, H.J. et al. 2002, ApJ, 569, 245
\reference{} Penton, S.V., Shull, J.M. \& Stocke, J.T. 2000, ApJ, 544, 150
\reference{} Pietz, J., Kerp, J., Kalberla, P.M.W., Mebold, U., Burton, W.B. \&
Hartmann, D. 1996, A\&A, 308, 37
\reference{}Putman, M.E. et al. 2002, AJ, 123, 873
\reference{}Putman, M.E. 2000, PhD Thesis, Australian National University
\reference{} Putman, M.E., Staveley-Smith, L., Gibson, B.K. et al. 1998,
Nature, 394, 752
\reference{} Quilis, V. \& Moore, B. 2001, ApJ, 555, L95
\reference{} Quilis, V., Moore, B. \& Bower, R. 2000 Science, 288, 5471, 1617
\reference{} Riediger, R., Petitjean, P. \& Mucket, J.P. 1998, A\&A, 329, 
30
\reference{} Rosenberg, J. \& Schneider, S. 2002, ApJ, 567, 247
\reference{} Ryder, S. et al. 2001, ApJ, 555, 232
\reference{} Schmoldt, I. \& Saha, P. 1998, AJ, 115, 2231
\reference{} Schweitzer, A. 1996, PhD Thesis, UW-Madison
\reference{} Smith, B.J. 2000, ApJ, 541, 624
\reference{} Stanimirovic, S., Staveley-Smith, L., Dickey, J.M., Sault, R.J. \& Snowden, S.L. 1999, MNRAS, 302, 417
\reference{} Staveley-Smith, L., Kim, S., Calabretta, M. R., Haynes, R.F. \&
Kesteven, M.J. 2002, \mnras, in preparation 
\reference{} Staveley-Smith, L.\etal\ 1996, PASA, 13, 243
\reference{} van der Marel, R.P. 2001, ApJ, 122, 1827
\reference{} Wakker, B.P. 2001, ApJS, 136, 463
\reference{} Wakker, B.P., Kalberla, P.M.W., van Woerden, H., de Boer,
K.S., \& Putman, M.E. 2001, ApJS, 136, 537
\reference{} Wakker, B.P. \& van Woerden, H. 1997, ARA\&A, 35, 217
\reference{} Wakker, B.P. \& Bregman, J. 1990, in Interstellar Neutral Hydrogen at High Velocities, PhD Thesis of BP Wakker,
Rijks Univ. Groningen, ch. 5

\reference{} Wannier, P. \& Wrixon, G. T. 1972, ApJL, 173, L119
\reference{} Wayte, S.R. 1989, PASA, 8, 195
\reference{} Weiner, B.J. \& Williams, T.B. 1996, AJ, 111, 1156
\reference{}Yoshizawa, A. 1998, PhD Thesis, Tohoku University, Sendai, Japan
\reference{} Yun, M.S., Ho. P.T.P. \& Lo K.Y. 1994, Nature, 372, 530
\reference{} Yusef-Zadeh, F., Wardle, M. \& Parastaran, P. 1997, ApJL, 475,
  L119
\reference{} Zaritsky, D. \& Gonzalez, A.H. 1999, \pasp, 111, 1508
\reference{} Zaritsky, D., Smith, R., Frenk, C. \& White, S.D.M. 1997, \apj,
  478, 39
\reference{} Zwaan, M.A. 2001, MNRAS, 325, 1142
\reference{} Zwaan, M.A., Verheijen, M. \& Briggs, F. 1999, PASA, 16, 100

\end{references}
\end{document}